\documentclass{emulateapj}
\usepackage{natbib}
\usepackage{epsfig}
\usepackage{graphicx}
\usepackage{subfigure}
\usepackage{float}
\usepackage{amsmath}
\usepackage{color}
\usepackage{amssymb}
\usepackage{amsfonts}
\usepackage{apjfonts}
\usepackage[colorlinks,linkcolor=blue,anchorcolor=green,citecolor=blue]{hyperref}
\shorttitle{Measuring the Cosmic Curvature}
\shortauthors{Wei \& Wu}
\begin{document}

\title{An Improved Method to Measure the Cosmic Curvature}

\author{Jun-Jie Wei\altaffilmark{1,2} and Xue-Feng Wu\altaffilmark{1,3,4}}
\affil{$^1$ Purple Mountain Observatory, Chinese Academy of Sciences, Nanjing 210008, China; jjwei@pmo.ac.cn \\
$^2$ Guangxi Key Laboratory for Relativistic Astrophysics, Nanning 530004, China \\
$^3$School of Astronomy and Space Science, University of Science and Technology of China, Hefei, Anhui 230026, China\\
$^4$ Joint Center for Particle, Nuclear Physics and Cosmology, Nanjing University-Purple Mountain Observatory, Nanjing 210008, China}

\begin{abstract}
In this paper, we propose an improved model-independent method to constrain
the cosmic curvature by combining the most recent Hubble parameter $H(z)$ and supernovae
Ia (SNe Ia) data. Based on the $H(z)$ data, we first use the model-independent smoothing
technique, Gaussian processes, to construct distance modulus $\mu_{H}(z)$, which is susceptible to
the cosmic curvature parameter $\Omega_{k}$. In contrary to previous studies, the
light-curve fitting parameters, which account for distance estimation of SN ($\mu_{SN}(z)$),
are set free to investigate whether $\Omega_{k}$ has a dependence on them. By comparing $\mu_{H}(z)$
to $\mu_{SN}(z)$, we put limits on $\Omega_{k}$. Our results confirm that $\Omega_{k}$
is independent of the SN light-curve parameters. Moreover, we show that the measured
$\Omega_{k}$ is in good agreement with zero cosmic curvature, implying that there is
no significant deviation from a flat Universe at the current observational data level.
We also test the influence of different $H(z)$ samples and different Hubble constant $H_{0}$ values,
finding that different $H(z)$ samples do not present significant impact on the constraints.
However, different $H_{0}$ priors can affect the constraints of $\Omega_{k}$ in some degree.
The prior of $H_{0}=73.24\pm1.74$ km $\rm s^{-1}$ $\rm Mpc^{-1}$ gives a value of $\Omega_{k}$
a little bit above $1\sigma$ confidence level away from 0, but
$H_{0}=69.6\pm0.7$ km $\rm s^{-1}$ $\rm Mpc^{-1}$ gives it below $1\sigma$.

\end{abstract}

\keywords{cosmological parameters --- cosmology: observations --- supernovae: general --- galaxies: general}

\section{Introduction}

The cosmic curvature is one of the fundamental parameters in modern cosmology.
The intriguing question of whether the cosmic space is open, flat, or closed is
closely related to many important problems such as the evolution of our Universe,
the nature of dark energy, etc. A significant detection of a nonzero curvature
will have far-reaching consequences for mankind's views of fundamental physics
and inflation theory, since a flat Universe is supported by most of the
observational data, including the latest Plank result \citep{2016A&A...594A..13P}.

However, as a result of the strong degeneracy between the
spatial curvature and the dark energy equation of state, it is fairly difficult to
constrain these two parameters simultaneously. The curvature parameter
is generally treated as zero in a dark energy analysis, or conversely, some specific models
of dark energy (e.g., the cosmological constant) are assumed when constraining
the curvature. It should be underlined that a simple flatness assumption may leads to incorrect
reconstruction in the equation of state of dark energy even if the real curvature
is tiny \citep{2007JCAP...08..011C}, and some confusions between the flat $\Lambda$CDM
model and a dynamical dark energy non-flat model may be caused by a cosmological constant
assumption \citep{2008JCAP...12..008V}. In order to overcome the defects of a zero curvature
assumption, a direct model-independent method for determining the curvature by combining
measurements of the angular diameter distance $D_{A}(z)$ (or the luminosity distance $D_{L}(z)$)
and the Hubble parameter $H(z)$ has been proposed \citep{2007JCAP...08..011C,2008PhRvL.101a1301C}:
\begin{equation}
\Omega_{k}=\frac{\left[H(z)D'(z)\right]^{2}-c^{2}}{\left[H_{0}D(z)\right]^{2}}\;,
\label{eq:clarkson}
\end{equation}
where $c$ is the speed of light, $H_{0}$ is the Hubble constant,
$D(z)=(1+z)D_{A}(z)=D_{L}(z)/(1+z)$ is the comoving angular diameter distance,
and $D'(z)={\rm d}D(z)/{\rm d}z$ represents the derivative with respect to
the redshift $z$.

Since this method was proposed, it has been used to determine the curvature parameter
in several instances, including the following representative cases:
\citet{2010PhRvD..81h3537S} used the luminosity distances derived from Type Ia supernovae
(SNe Ia) observations, Hubble rate measurements inferred from passively evolving galaxies
and from baryon acoustic oscillation (BAO) data, and found no evidence for deviation from flatness
(see also~\citealt{2011arXiv1102.4485M});
\citet{2014PhRvD..90b3012S} compared four different measurement techniques to test the cosmic curvature
from the most recent Hubble rate and SNe Ia data; \citet{2014ApJ...789L..15L} determined the curvature
parameter by using $H(z)$ and $D_{A}(z)$ pairs from BAO measurements;
\citet{2016PhRvD..93d3517C} used the model-independent smoothing technique (i.e, Gaussian process) to
reconstruct $H(z)$ from differential ages of galaxies and from radial BAO data and $D_{L}(z)$ from
the SNe Ia Union2.1 data sets and then measure the curvature; \citet{2017JCAP...01..015L} tested
the flatness of the Universe at redshifts 0.32 and 0.57 using the  most recent BAO and SNe Ia data,
and they found that the current observations are compatible with a flat Universe;
and \citet{2016ApJ...828...85Y} constrained the curvature to be $\Omega_{k}=-0.09\pm0.19$, combining
the measurements of $H(z)$ derived from differential ages of galaxies and from radial BAO
data with $D_{A}(z)$ estimated from BAO data.

In principle, the nuisance parameters characterizing SN light-curves should be optimized simultaneously
with the cosmological parameters when using SNe Ia as standard candles. But it is shown that
the nuisance parameters have extremely little covariance with the cosmological parameters (see~\citealt{2011ApJ...740...72M}).
In previous works, the luminosity distances of SNe Ia were obtained directly from Hubble diagrams where
the light-curve fitting (nuisance) parameters were inferred from global fitting within the context of a cosmological
model. To confirm if the cosmic curvature parameter has a dependence on the nuisance parameters or not, we keep them free in our analysis.
On the other hand, in the method of \citet{2007JCAP...08..011C,2008PhRvL.101a1301C}, one needs to estimate
the derivative function of $D(z)$ from a fitting function (see Equation~(\ref{eq:clarkson})), which will introduce
a large uncertainty \citep{2016ApJ...828...85Y}.
In order to avoid the shortcoming of this method, we perform an improved
model-independent method to achieve a reasonable and compelling test
of the cosmic curvature. Moreover, we also investigate the impact of Hubble constant
$H_{0}$ on this test.

The rest of this paper is arranged as follows.
In Section~\ref{sec:data}, we briefly describe the data used in our work, including
the most recent SNe Ia and $H(z)$ data. In Section~\ref{sec:method}, we introduce our improved
method for testing the curvature. The constraints on the curvature are shown in
Section~\ref{sec:result}. Finally, we summarize our conclusions in Section~\ref{sec:summary}.

\section{Observational data}\label{sec:data}
In the following, we describe the data sets that we will use in the present analysis.

\subsection{SNe Ia sample}
We use a joint light-curve analysis (JLA) sample of 740 SNe Ia processed by
\citet{2014A&A...568A..22B}. The observed distance modulus of each SN is given by
\begin{equation}
\mu_{\rm SN}=m^{\star}_{B}+\alpha\cdot X_{1}-\beta\cdot \mathcal{C}-M_{B}\;,
\label{eq:SNmu}
\end{equation}
where $m^{\star}_{B}$ is the observed peak magnitude in rest frame $B$ band, $X_{1}$
describes the time stretching of light-curve, and $\mathcal{C}$ corresponds to the
supernova color at maximum brightness. The absolute $B$-band magnitude $M_{B}$
is assumed to be related to the host stellar mass ($M_{\rm stellar}$) by
a simple step function \citep{2014A&A...568A..22B}:
\begin{equation}\label{HSFR}
  M_{B} = \left\lbrace \begin{array}{ll} M^{1}_{B}~~~~~~~~~~~~~~~{\rm for}~~~M_{\rm stellar}<10^{10}M_{\odot},\\
                                         M^{1}_{B}+\Delta_{M}~~~~~{\rm otherwise}. \\
\end{array} \right.
\end{equation}
Notice that $\alpha$, $\beta$, $M^{1}_{B}$, and $\Delta_{M}$ are nuisance parameters in the
distance estimate, which should be fitted simultaneously with the cosmological parameters.
While, $m^{\star}_{B}$, $X_{1}$, and $\mathcal{C}$ are obtained from the observed
SN light-curve.

For each SN, the theoretical distance modulus $\mu_{\rm th}$ can be calculated from
the measured redshift $z$ by the definition:
\begin{equation}
\mu_{\rm th}\equiv5\log\left[\frac{D_{L}(z)}{\rm Mpc}\right]+25\;,
\label{eq:mu}
\end{equation}
where $D_{L}(z)$ is the cosmology-dependent luminosity distance.
\citet{2014A&A...568A..22B} fit a $\Lambda$CDM cosmology to the JLA sample by minimizing
the $\chi^{2}$ statistic:
\begin{equation}
\chi^{2}=\bf{\Delta \hat{\mu}}^{T}\cdot \textbf{Cov}^{-1}\cdot \bf{\Delta \hat{\mu}}\;,
\end{equation}
where $\Delta \hat{\mu}=\hat{\mu}_{\rm SN}(\alpha,\;\beta,\;M^{1}_{B},\;\Delta_{M};\;z)
-\hat{\mu}^{\rm \Lambda CDM}_{\rm th}(\Omega_{m},\;H_{0};\;z)$ is the data vector
and \textbf{Cov} is the full covariance matrix, defined by
\begin{equation}
\textbf{Cov}=\textbf{D}_{\rm stat}+\textbf{C}_{\rm stat}+\textbf{C}_{\rm sys}\;.
\end{equation}
Here $\textbf{D}_{\rm stat}$ is the diagonal part of the statistical uncertainty, given by
\begin{equation}
\begin{split}
(\textbf{D}_{\rm stat})_{ii}=\sigma^{2}_{m_{B},i}+\alpha^{2}\sigma^{2}_{X_{1},i}+\beta^{2}\sigma^{2}_{\mathcal{C},i}\qquad\qquad\qquad~~~~~\\
+2\alpha C_{m_{B}\,X_{1},\,i}-2\beta C_{m_{B}\,\mathcal{C},\,i}-2\alpha\beta C_{X_{1}\,\mathcal{C},\,i}\qquad\\
+\sigma^{2}_{\rm lens}+\left(\frac{5\sigma_{z,i}}{z_{i}\ln 10}\right)^{2}+\sigma^{2}_{\rm int}\;,\qquad\qquad~~~~~~~~
\end{split}
\label{eq:dstat}
\end{equation}
where the last three terms stand for the variation of magnitudes arisen from gravitational lensing,
the uncertainty in cosmological redshift caused by peculiar velocities, and the intrinsic variation
in SN magnitude, respectively. $\sigma_{{m_{B}},i}$, $\sigma_{X_{1},i}$, and
$\sigma_{{\mathcal{C}},i}$ represent the standard errors of the peak magnitude
and light-curve parameters of the $i$-th~SN\null. The terms $C_{m_{B}\,X_{1},\,i},\;C_{m_{B}\,\mathcal{C},\,i}$,
and $C_{X_{1}\,\mathcal{C},\,i}$ denote the covariances among $m_{B},\;X_{1},\;\mathcal{C}$
for the $i$-th~SN\null.
The statistical and systematic covariance matrices,
$\textbf{C}_{\rm stat}$ and $\textbf{C}_{\rm sys}$, are given by
\begin{equation}
\textbf{C}_{\rm stat}+\textbf{C}_{\rm sys}=V_{0}+\alpha^{2}V_{a}+\beta^{2}V_{b}
+2\alpha V_{0a}-2\beta V_{0b}-2\alpha \beta V_{ab}\;,
\label{eq:cov}
\end{equation}
where $V_{0}$, $V_{a}$, $V_{b}$, $V_{0a}$, $V_{0b}$, and $V_{ab}$ are matrices
available in \citet{2014A&A...568A..22B}.
Since the Hubble constant $H_{0}$ is degenerate with $M_{B}$ when constructing
an SN Hubble diagram, it is not free if $M_{B}$ is considered as one of the optimized variables.
\citet{2014A&A...568A..22B} fixed the value of $H_{0}=70$ km $\rm s^{-1}$ $\rm Mpc^{-1}$,
and they obtained $(\alpha,\beta,M^{1}_{B},\Delta_{M})=(0.141\pm0.006,3.101\pm0.075,-19.05\pm0.02,
-0.070\pm0.023)$ including both statistical and systematic errors.

In this work, we directly adopt the observational
quantities ($m^{\star}_{B}$, $X_{1}$, $\mathcal{C}$) from the JLA sample
to constrain the curvature. By marginalizing the nuisance parameters
($\alpha$, $\beta$, $M^{1}_{B}$, $\Delta_{M}$), one can obtain a cosmology-independent
constraint on the curvature and justify whether the curvature has a dependence on the nuisance parameters.

\subsection{Hubble parameter data}
The $H(z)$ measurement can be obtained via two ways. One is calculating the differential ages
of passively evolving galaxies (e.g.,~\citealt{2002ApJ...573...37J,2005PhRvD..71l3001S,2010JCAP...02..008S}),
usually called cosmic chronometer (hereafter CC $H(z)$). The other is based on the detection of radial BAO
features (e.g.,~\citealt{2009MNRAS.399.1663G,2012MNRAS.425..405B,2013MNRAS.429.1514S}). For convenience,
we name this kind of $H(z)$ as BAO $H(z)$. We compile the latest 41 $H(z)$ data points in Table~\ref{table1},
including 31 CC $H(z)$ data and 10 BAO $H(z)$ data. These are all independent datasets and analyses.

\begin{table}
\centering \caption{The latest $H(z)$ measurements from the differential age method (I)
and the radial BAO method (II)}
\begin{tabular}{cccc}
\hline
\hline
 \emph{z} & $H(z)$ & Method & References \\
 & (km $\rm s^{-1}$ $\rm Mpc^{-1}$) &  & \\
\hline
0.09	&	$	69	\pm	12	$	& I &	\citet{2003ApJ...593..622J} \\
\hline
0.17	&	$	83	\pm	8	$	&I &	\\
0.27	&	$	77	\pm	14	$	&I &	\\
0.4	&	$	95	\pm	17	$	&I &	 \\
0.9	&	$	117	\pm	23	$	&I &	\citet{2005PhRvD..71l3001S} \\
1.3	&	$	168	\pm	17	$	&I &	\\
1.43	&	$	177	\pm	18	$	&I &	\\
1.53	&	$	140	\pm	14	$	&I &	\\
1.75	&	$	202	\pm	40	$	&I &	\\
\hline
0.48	&	$	97	\pm	62	$	&I &	\citet{2010JCAP...02..008S} \\
0.88	&	$	90	\pm	40	$	&I &	\\
\hline
0.35	&	$	82.1	\pm	4.9	$	&I &	\citet{2012MNRAS.426..226C} \\
\hline
0.179	&	$	75	\pm	4	$	&I &	\\
0.199	&	$	75	\pm	5	$	&I &	\\
0.352	&	$	83	\pm	14	$	&I &	\\
0.593	&	$	104	\pm	13	$	&I &	\citet{2012JCAP...07..053M} \\
0.68	&	$	92	\pm	8	$	&I &	\\
0.781	&	$	105	\pm	12	$	&I &	\\
0.875	&	$	125	\pm	17	$	&I &	\\
1.037	&	$	154	\pm	20	$	&I &	\\
\hline
0.07	&	$	69	\pm	19.6	$	&I &	\\
0.12	&	$	68.6	\pm	26.2	$	&I &	\citet{2014RAA....14.1221Z} \\
0.2	&	$	72.9	\pm	29.6	$	&I &	\\
0.28	&	$	88.8	\pm	36.6	$	&I &	\\
\hline
1.363	&	$	160	\pm	33.6	$	&I &	\citet{2015MNRAS.450L..16M} \\
1.965	&	$	186.5	\pm	50.4	$	&I &	\\
\hline
0.3802	&	$	83	\pm	13.5	$	&I &	\\
0.4004	&	$	77	\pm	10.2	$	&I &	\\
0.4247	&	$	87.1	\pm	11.2	$	&I &	\citet{2016JCAP...05..014M} \\
0.4497	&	$	92.8	\pm	12.9	$	&I &	\\
0.4783	&	$	80.9	\pm	9	$	&I &	\\
\hline
0.24	&	$	79.69	\pm	2.65	$	&II &	\citet{2009MNRAS.399.1663G} \\
0.43	&	$	86.45	\pm	3.68	$	&II &	\\
\hline
0.44	&	$	82.6	\pm	7.8	$	&II &	\\
0.6	&	$	87.9	\pm	6.1	$	&II &	\citet{2012MNRAS.425..405B}\\
0.73	&	$	97.3	\pm	7	$	&II &	\\
\hline
0.35	&	$	84.4	\pm	7	$	&II &	\citet{2013MNRAS.431.2834X}\\
\hline
0.57	&	$	92.4	\pm	4.5	$	&II &	\citet{2013MNRAS.429.1514S}\\
\hline
2.3	&	$	224	\pm	8	$	&II &	 \citet{Busca2013}\\
\hline
2.36	&	$	226	\pm	8	$	&II &	\citet{2014JCAP...05..027F}\\
\hline
2.34	&	$	222	\pm	7	$	&II &	\citet{Delubac2015} \\
\hline
\end{tabular}
\label{table1}
\end{table}

\section{New Model-independent Method}\label{sec:method}

Within the framework of Friedmann-Robertson-Walker metric, the proper distance
can be written as \citep{1999astro.ph..5116H}
\begin{equation}
d_{P}(z)=\frac{c}{H_{0}}\int^{z}_{0}\frac{dz'}{E(z')}\;,
\label{eq:dp}
\end{equation}
where $E(z)=H(z)/H_{0}$. For the base $\Lambda$CDM model,
$E(z)$ has the form of
$E(z)=\sqrt{\Omega_{m}(1+z)^{3}+\Omega_{\Lambda}+\Omega_{k}(1+z)^{2}}$.

\begin{figure*}
\begin{tabular}{c}
\includegraphics[keepaspectratio,clip,width=1.0\textwidth]{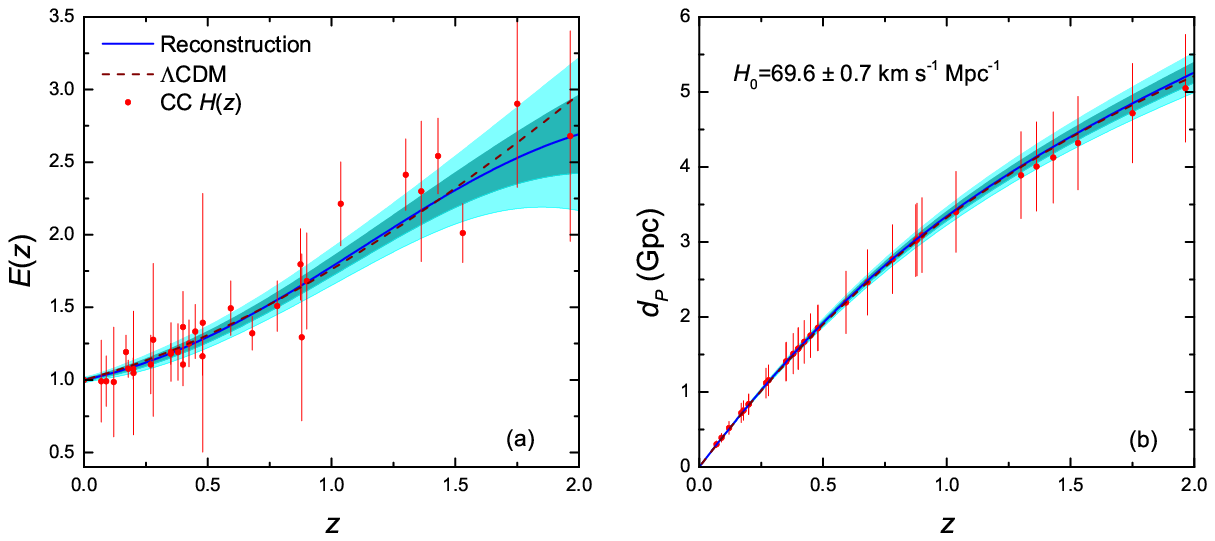} \\
\includegraphics[keepaspectratio,clip,width=1.0\textwidth]{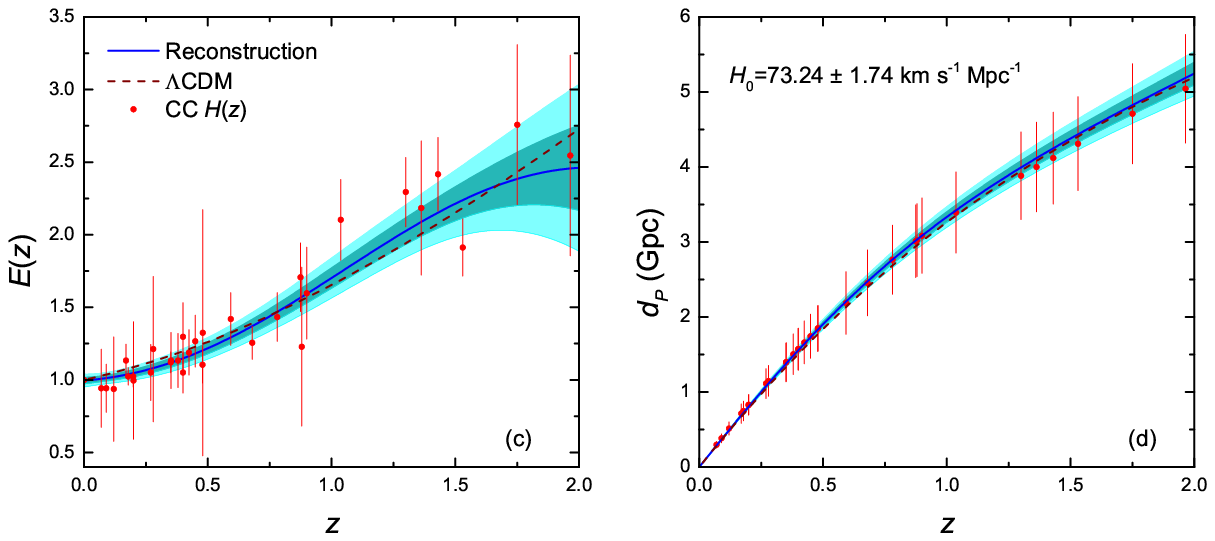}
\end{tabular}
\vskip-0.1in
\caption{{\it Top row}: Gaussian process reconstruction of $E(z)$ (panel (a); solid line) for the CC $H(z)$ data
with $H_{0}=69.6\pm0.7$ km $\rm s^{-1}$ $\rm Mpc^{-1}$. The observed (red points) and reconstructed
(solid lines) $d_{P}(z)$ in panel (b) are derived from the observations and reconstructions of $E(z)$, respectively.
The shaded regions are the $1\sigma$ and $2\sigma$ confidence regions of the reconstruction.
{\it Bottom row}: Same as {\it Top row}, except now for the CC $H(z)$ data with $H_{0}=73.24\pm1.74$ km $\rm s^{-1}$ $\rm Mpc^{-1}$.
The dashed lines correspond to the best-fit flat $\Lambda$CDM models with $\Omega_{m}=0.30$ ({\it Top row})
and $\Omega_{m}=0.25$ ({\it Bottom row}), respectively.}
\label{f1}
\end{figure*}

\begin{figure*}
\begin{tabular}{c}
\includegraphics[keepaspectratio,clip,width=1.0\textwidth]{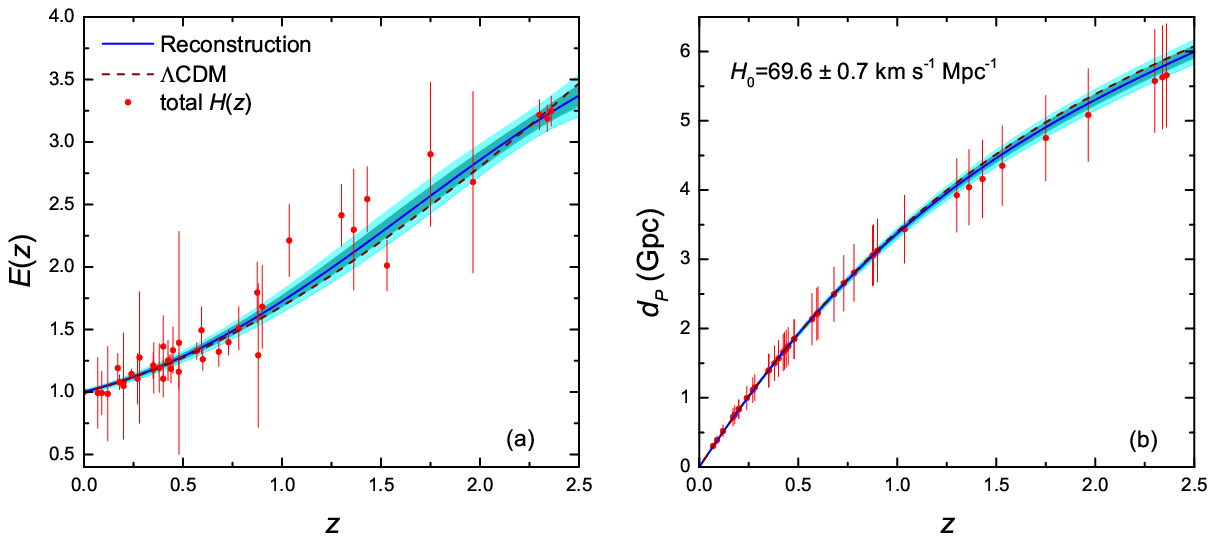} \\
\includegraphics[keepaspectratio,clip,width=1.0\textwidth]{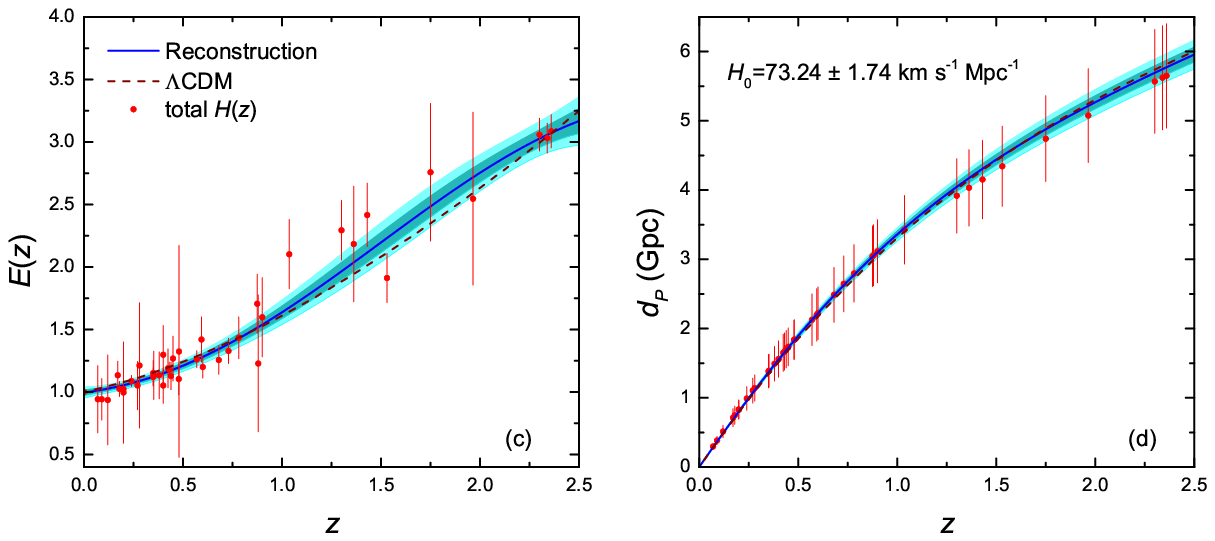}
\end{tabular}
\vskip-0.1in
\caption{Same as Figure~\ref{f1}, except now for the total $H(z)$ data.
In this case, the dashed lines correspond to the best-fit flat $\Lambda$CDM models with $\Omega_{m}=0.26$ ({\it Top row})
and $\Omega_{m}=0.23$ ({\it Bottom row}), respectively.}
\label{f2}
\end{figure*}

Inspired by the work of \citet{2016ApJ...828...85Y}, we employ
an improved approach to acquire proper distances that are independent of any
specific cosmological model. The detailed procedures of our approach are described
as follows:

\begin{enumerate}

\item Since the proper distance $d_{P}$ only depends on the $E(z)$ function, one can
reconstruct the model-independent $E(z)$ function from the $H(z)$ measurements
and then derive $d_{P}$ with Equation~(\ref{eq:dp}).

\item We use the model-independent method Gaussian processes (GP) to reconstruct $E(z)$.
GP allow one to reconstruct a function from data directly without any parametric
assumption. In this process, the reconstructed function $f(z)$ at different points
$z$ and $\tilde{z}$ are correlated by a covariance function $k(z,\tilde{z})$,
which only depends on two hyperparameters $l$ and $\sigma_{f}$. Both $l$ and $\sigma_{f}$
would be determined by GP with the observational data. Therefore, the GP method does not specify
any form of $f(z)$ and is model-independent. There is a python package
of GP developed by \citet{2012JCAP...06..036S}, which has been widely used in various studies
(e.g.,~\citealt{2012MNRAS.425.1664B,2012PhRvD..86h3001S,2012PhRvD..85l3530S,2013arXiv1311.6678S,
2014PhRvD..89b3503Y,2014MNRAS.441L..11B,2015PhRvD..91l3533Y,2016PhRvD..93d3517C,2016ApJ...828...85Y,2016JCAP...12..005Z}).
We refer the reader to \citet{2012JCAP...06..036S} for more details on the GP method and the GP code.

\item We normalize the $H(z)$ data using an independent measurement of the local Hubble parameter $H_{0}$;
thus we get the dimensionless Hubble parameter $E(z)=H(z)/H_{0}$. Note that the initial condition $E(z=0)=1$
should be taken into account in our calculation. Considering the uncertainty of Hubble constant, the propagated error of $E(z)$
can be calculated by $\sigma^{2}_{E}=\left(\sigma^{2}_{H}/H^{2}_{0}\right)+\left(H^{2}/H^{4}_{0}\right)\sigma^{2}_{H_{0}}$.
To explore the influence of Hubble constant on the reconstruction and then on the test of
the curvature parameter (more on this below), we follow the treatment of \citet{2016JCAP...12..005Z}
and adopt two recent measurements $H_{0}=69.6\pm0.7$ km $\rm s^{-1}$ $\rm Mpc^{-1}$ with $1\%$
uncertainty \citep{2014ApJ...794..135B} and $H_{0}=73.24\pm1.74$ km $\rm s^{-1}$ $\rm Mpc^{-1}$
with $2.4\%$ uncertainty \citep{2016ApJ...826...56R}, respectively.
Moreover, we study the potential impact on the results from different $H(z)$ samples
(i.e., the only CC $H(z)$ data and the total $H(z)$ data). We show the results in Figures~\ref{f1}
and \ref{f2}.

\end{enumerate}

Using the GP method, the reconstructions of $E(z)$ for the CC $H(z)$ data with $H_{0}=69.6\pm0.7$ km $\rm s^{-1}$ $\rm Mpc^{-1}$
and with $H_{0}=73.24\pm1.74$ km $\rm s^{-1}$ $\rm Mpc^{-1}$ are shown in Figures~\ref{f1}(a) and~\ref{f1}(c), respectively.
The solid lines are the means of the reconstructions and the shaded regions are the $1\sigma$
and $2\sigma$ confidence regions of the reconstructions. Because of the poor quality of data
at higher redshifts, the errors become larger. For a comparison, we also fit the observational
data points of $E(z)$ using the flat $\Lambda$CDM model (dashed lines). One can see from these plots
that the reconstructions of $E(z)$ are well consistent with the best-fit flat $\Lambda$CDM model
within their $1\sigma$ confidence regions, indicating that the GP method can give a reliable
reconstructed function from the observational data. With the observations and reconstructions
of $E(z)$, we can use Equation~(\ref{eq:dp}) to derive the observed $d_{P}(z)$ together with their
$1\sigma$ errors and the reconstructed $d_{P}(z)$ together with the $1\sigma$ and $2\sigma$ confidence
levels at a certain $z$, respectively. As shown in Figures~\ref{f1}(b) and~\ref{f1}(d), both
the observed (red points) and reconstructed (solid lines) $d_{P}(z)$ are also consistent
with those determined from the best-fit flat $\Lambda$CDM model (dashed lines).
Not surprisingly, the comparison between the top and bottom panels in Figure~\ref{f1} shows that
the best-fit values of $\Omega_{m}$ for the flat $\Lambda$CDM model are different, since different $H_{0}$ priors are adopted.

We follow the same procedure for the total $H(z)$ data, first considering a prior of
$H_{0}=69.6\pm0.7$ km $\rm s^{-1}$ $\rm Mpc^{-1}$ (the top row of Figure~\ref{f2}), followed by the other one of
$H_{0}=73.24\pm1.74$ km $\rm s^{-1}$ $\rm Mpc^{-1}$ (the bottom row of Figure~\ref{f2}).
The comparison between these two $H(z)$ samples may be summarized as follows: the reconstructions
of $E(z)$ and $d_{P}(z)$ are well consistent with the flat $\Lambda$CDM model for both the CC $H(z)$
and total $H(z)$ data, suggesting that the GP method can reconstruct the $E(z)$ and $d_{P}(z)$ functions well;
the errors of the reconstructions for the total $H(z)$ data become smaller owing to the added BAO $H(z)$ data.

By using the reconstructed $d_{P}(z)$ function together with its $1\sigma$ uncertainty $\sigma_{d_{P}}$,
the luminosity distance $D^{H}_{L}$ from the $H(z)$ data can be expressed as
\begin{equation}\label{HSFR}
\frac{D^{H}_{L}(z)}{(1+z)} = \left\lbrace \begin{array}{lll} \frac{c}{H_{0}}\frac{1}{\sqrt{|\Omega_{k}|}}\sinh\left[\sqrt{|\Omega_{k}|}d_{P}(z)\frac{H_{0}}{c}\right]~~{\rm for}~~\Omega_{k}>0\\
                                         d_{P}(z)~~~~~~~~~~~~~~~~~~~~~~~~~~~~~~~~~~~{\rm for}~~\Omega_{k}=0\;, \\
                                         \frac{c}{H_{0}}\frac{1}{\sqrt{|\Omega_{k}|}}\sin\left[\sqrt{|\Omega_{k}|}d_{P}(z)\frac{H_{0}}{c}\right]~~~~{\rm for}~~\Omega_{k}<0\\
\end{array} \right.
\end{equation}
with its corresponding uncertainty
\begin{equation}
\sigma_{D^{H}_{L}} = \left\lbrace \begin{array}{lll} (1+z)\cosh\left[\sqrt{|\Omega_{k}|}d_{P}(z)\frac{H_{0}}{c}\right]\sigma_{d_{P}}~~{\rm for}~~\Omega_{k}>0\\
                                         (1+z)\sigma_{d_{P}}~~~~~~~~~~~~~~~~~~~~~~~~~~~~~~~{\rm for}~~\Omega_{k}=0\;, \\
                                         (1+z)\cos\left[\sqrt{|\Omega_{k}|}d_{P}(z)\frac{H_{0}}{c}\right]\sigma_{d_{P}}~~~~{\rm for}~~\Omega_{k}<0\\
\end{array} \right.
\end{equation}
where we emphasize that the spatial curvature $\Omega_{k}$ is the only one free parameter.
Then, we can further obtain the reconstructed distance modulus $\mu_{H}(\Omega_{k};\,z)$ from the $H(z)$ data by
\begin{equation}
\mu_{H}(\Omega_{k};\,z)=5\log\left[\frac{D^{H}_{L}(\Omega_{k};\,z)}{\rm Mpc}\right]+25\;.
\end{equation}
The propagated uncertainty of $\mu_{H}(\Omega_{k};\,z)$ is given by
\begin{equation}
\sigma_{\mu_{H}}=\frac{5}{\ln10}\frac{\sigma_{D^{H}_{L}}}{D^{H}_{L}}\;.
\end{equation}

Now, we use a $\chi^{2}$ minimization to constrain $\Omega_{k}$,
\begin{equation}
\chi^{2}(\alpha,\;\beta,\;M^{1}_{B},\;\Delta_{M},\;\Omega_{k})
=\bf{\Delta \hat{\mu}}^{T}\cdot \textbf{Cov}^{-1}\cdot \bf{\Delta \hat{\mu}}\;,
\end{equation}
where $\Delta \hat{\mu}=\hat{\mu}_{\rm SN}(\alpha,\;\beta,\;M^{1}_{B},\;\Delta_{M};\;z)
-\hat{\mu}_{H}(\Omega_{k};\;z)$ is the difference between the distance moduli
$\mu_{\rm SN}$ of SNe Ia derived from Equation~(\ref{eq:SNmu}) and the constructed distance moduli
$\mu_{H}$ from the $H(z)$ data, and $\bf{Cov}=\bf{\bar{D}}_{\rm stat}+\bf{C}_{\rm stat}+\bf{C}_{\rm sys}$
is the full covariance matrix.
Here $\bf{\bar{D}}_{\rm stat}$ is the diagonal part of the statistical uncertainty, given by
\begin{equation}
(\bf{\bar{D}}_{\rm stat})_{\emph{ii}}=(\bf{D}_{\rm stat}^{SN})_{\emph{ii}}+\sigma^{2}_{\mu_{\emph{H}},\emph{i}}\;,
\end{equation}
where $\bf{D}_{\rm stat}^{SN}$ of SNe Ia comes from Equation~(\ref{eq:dstat}).
The statistical and systematic covariance matrices,
$\textbf{C}_{\rm stat}$ and $\textbf{C}_{\rm sys}$, are given by Equation~(\ref{eq:cov}).

The likelihood distributions of free parameters can be obtained by
$L(\alpha,\;\beta,\;M^{1}_{B},\;\Delta_{M},\;\Omega_{k})\propto \exp(-\chi^{2}/2)$.
We use the Markov Chain Monte Carlo technique to generate sample points distributed
in parameter space according to the posterior probability, using the Metropolis-Hastings algorithm
with uniform prior distributions. Then we apply a public python package ``triangle.py''
from \citet{2013PASP..125..306F} to plot our constraint contours.

\begin{table*}
\centering \caption{Best-fit values with $1\sigma$ and $2\sigma$ standard errors for
the cosmic curvature $\Omega_{k}$ and the SN nuisance parameters $(\alpha,\;\beta,\;M^{1}_{B},\;\Delta_{M})$}
\begin{tabular}{lcc|cc}
\hline
\hline
 & \multicolumn{2}{c|}{CC $H(z)$ + SNe Ia}  & \multicolumn{2}{c}{total $H(z)$ + SNe Ia} \\
\cline{2-5}
   & $H_{0}=69.6\pm0.7$  & $H_{0}=73.24\pm1.74$ & $H_{0}=69.6\pm0.7$  & $H_{0}=73.24\pm1.74$ \\
 & (km $\rm s^{-1}$ $\rm Mpc^{-1}$) &  (km $\rm s^{-1}$ $\rm Mpc^{-1}$) & (km $\rm s^{-1}$ $\rm Mpc^{-1}$) &  (km $\rm s^{-1}$ $\rm Mpc^{-1}$)\\
\hline
$\Omega_{k}$	&	$0.09\pm0.25(1\sigma)\pm0.49(2\sigma)$	    &	$-0.28\pm0.22(1\sigma)\pm0.43(2\sigma)$  &	 $-0.02\pm0.24(1\sigma)\pm0.47(2\sigma)$	    &	 $-0.35\pm0.22(1\sigma)\pm0.43(2\sigma)$ \\
$\alpha$	    &	$0.14\pm0.01(1\sigma)\pm0.02(2\sigma)$	    &	$0.14\pm0.01(1\sigma)\pm0.02(2\sigma)$   &	$0.14\pm0.01(1\sigma)\pm0.02(2\sigma)$	    &	 $0.14\pm0.01(1\sigma)\pm0.02(2\sigma)$ \\
$\beta$	        &	$3.10\pm0.09(1\sigma)\pm0.18(2\sigma)$	    &	$3.11\pm0.09(1\sigma)\pm0.18(2\sigma)$   &	$3.10\pm0.09(1\sigma)\pm0.18(2\sigma)$	    &	 $3.11\pm0.09(1\sigma)\pm0.18(2\sigma)$ \\
$M^{1}_{B}$	    &	$-19.08\pm0.02(1\sigma)\pm0.04(2\sigma)$	&	$-19.01\pm0.02(1\sigma)\pm0.04(2\sigma)$ &	 $-19.07\pm0.02(1\sigma)\pm0.04(2\sigma)$	&	 $-19.00\pm0.02(1\sigma)\pm0.04(2\sigma)$ \\
$\Delta_{M}$	&	$-0.07\pm0.03(1\sigma)\pm0.06(2\sigma)$	    &	$-0.07\pm0.03(1\sigma)\pm0.06(2\sigma)$  &	 $-0.07\pm0.03(1\sigma)\pm0.06(2\sigma)$	    &	 $-0.07\pm0.03(1\sigma)\pm0.06(2\sigma)$ \\
\hline
\end{tabular}
\label{table2}
\end{table*}

\section{Constraints on the cosmic curvature}\label{sec:result}
To investigate the influence of Hubble constant $H_{0}$ on the reconstruction of $E(z)$
and then on the test of the curvature parameter, we take into account two priors of $H_{0}$.
We also compare the tests from different $H(z)$ samples
(i.e., the only CC $H(z)$ data and the total $H(z)$ data).

\begin{figure*}
\centering
\vskip-0.1in
\begin{tabular}{c}
\includegraphics[keepaspectratio,clip,width=0.65\textwidth]{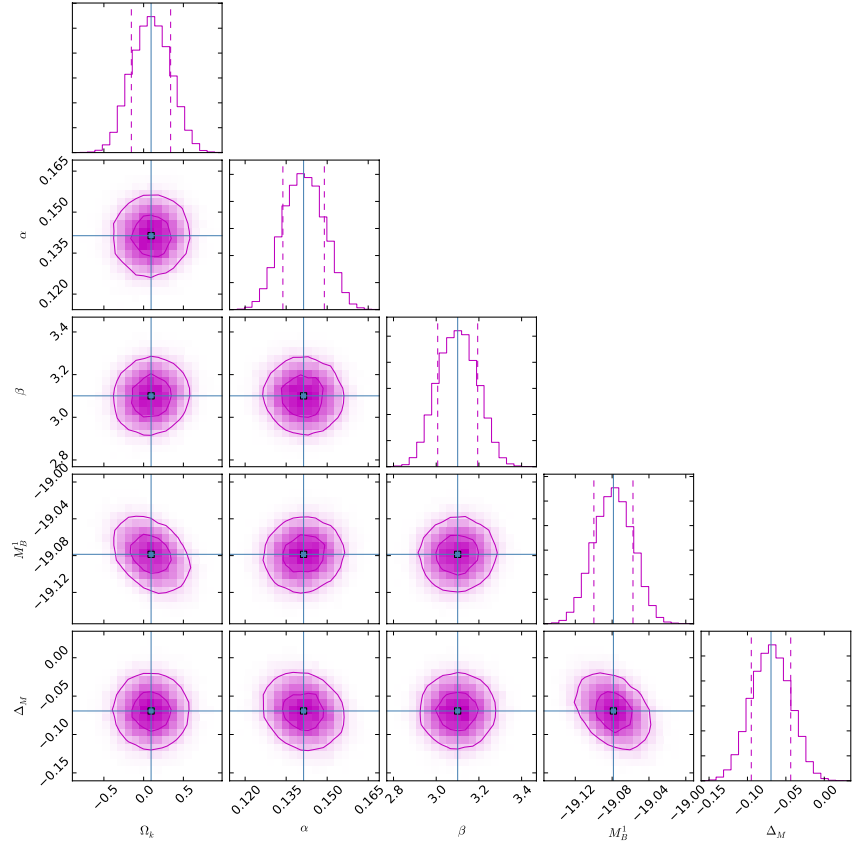} \\{(a) \;$H_{0}=69.6\pm0.7$ km $\rm s^{-1}$ $\rm Mpc^{-1}$}\\
\includegraphics[keepaspectratio,clip,width=0.65\textwidth]{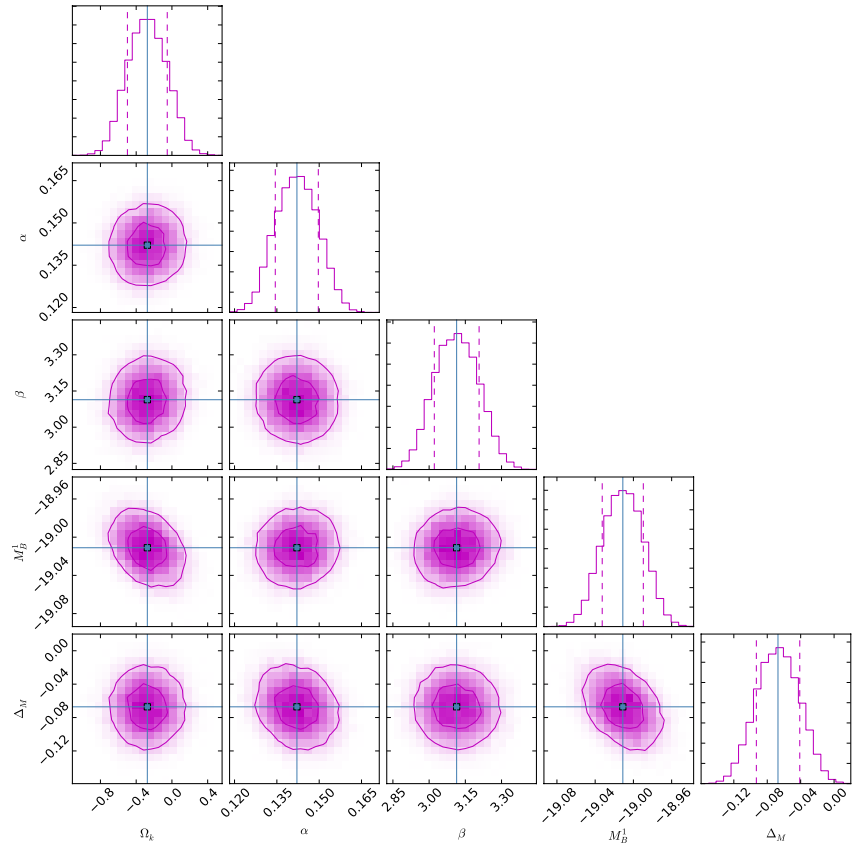} \\{(b) \;$H_{0}=73.24\pm1.74$ km $\rm s^{-1}$ $\rm Mpc^{-1}$}
\end{tabular}
\caption{(a): One-D marginalized distributions and two-D joint distributions with the $1\sigma$ and $2\sigma$
contours corresponding to the cosmic curvature $\Omega_{k}$ and the SNe Ia nuisance parameters $(\alpha,\;\beta,\;M^{1}_{B},\;\Delta_{M})$,
using the CC $H(z)$ + SNe Ia data with the prior of $H_{0}=69.6\pm0.7$ km $\rm s^{-1}$ $\rm Mpc^{-1}$.
The vertical solid lines denote the best-fits, and the vertical dashed lines enclose the $1\sigma$ confidence region.
(b): Same as panel (a), but now with the prior of $H_{0}=73.24\pm1.74$ km $\rm s^{-1}$ $\rm Mpc^{-1}$.}
\label{f3}
\end{figure*}

Applying the above $\chi^{2}$-minimization procedure, we find that the best-fit
curvature parameter using the CC $H(z)$ + JLA SNe Ia data with the prior of
$H_{0}=69.6\pm0.7$ km $\rm s^{-1}$ $\rm Mpc^{-1}$ is $\Omega_{k}=0.09\pm0.25(1\sigma)\pm0.49(2\sigma)$.
For the case of $H_{0}=73.24\pm1.74$ km $\rm s^{-1}$ $\rm Mpc^{-1}$,
we obtain $\Omega_{k}=-0.28\pm0.22(1\sigma)\pm0.43(2\sigma)$.
Our constraint results with these two $H_{0}$ priors are presented in Figures~\ref{f3}(a)
and~\ref{f3}(b), respectively. We give the 1-D distributions for each parameter
$(\Omega_{k},\;\alpha,\;\beta,\;M^{1}_{B},\;\Delta_{M})$, and $1\sigma$, $2\sigma$
contours for the joint distributions of any two parameters. The corresponding best-fit parameters
are summarized in Table~\ref{table2}, along with the $1\sigma$ and $2\sigma$ standard deviations for each.
From Figures~\ref{f3}(a) and~\ref{f3}(b), one can easily see that the measured $\Omega_{k}$ is consistent
with zero cosmic curvature within the $1.3\sigma$ confidence level for both of the two $H_{0}$ priors,
implying that there is no significant deviation from a flat Universe at the current observational
data [$H(z)$ data and SNe Ia] level.\footnote{Similar estimation of $\Omega_{k}$ from the $H(z)$ + SNe Ia data
was given in \cite{2016ApJ...833..240L}, which we received while working on this paper.}
However, a careful comparison of Figure~\ref{f3}(a) and
Figure~\ref{f3}(b) shows that different $H_{0}$ priors can affect the constraints on $\Omega_{k}$
in some degree.
The prior of $H_{0}=73.24\pm1.74$ km $\rm s^{-1}$ $\rm Mpc^{-1}$ gives a value of $\Omega_{k}$
a little bit above $1\sigma$ away from 0, but $H_{0}=69.6\pm0.7$ km $\rm s^{-1}$ $\rm Mpc^{-1}$
gives it below $1\sigma$. Note that SNe Ia data do not constrain $H_{0}$, so these different pulls
on $H_{0}$ are coming from the $H(z)$ constraints.

\begin{figure*}
\centering
\vskip-0.1in
\begin{tabular}{c}
\includegraphics[keepaspectratio,clip,width=0.65\textwidth]{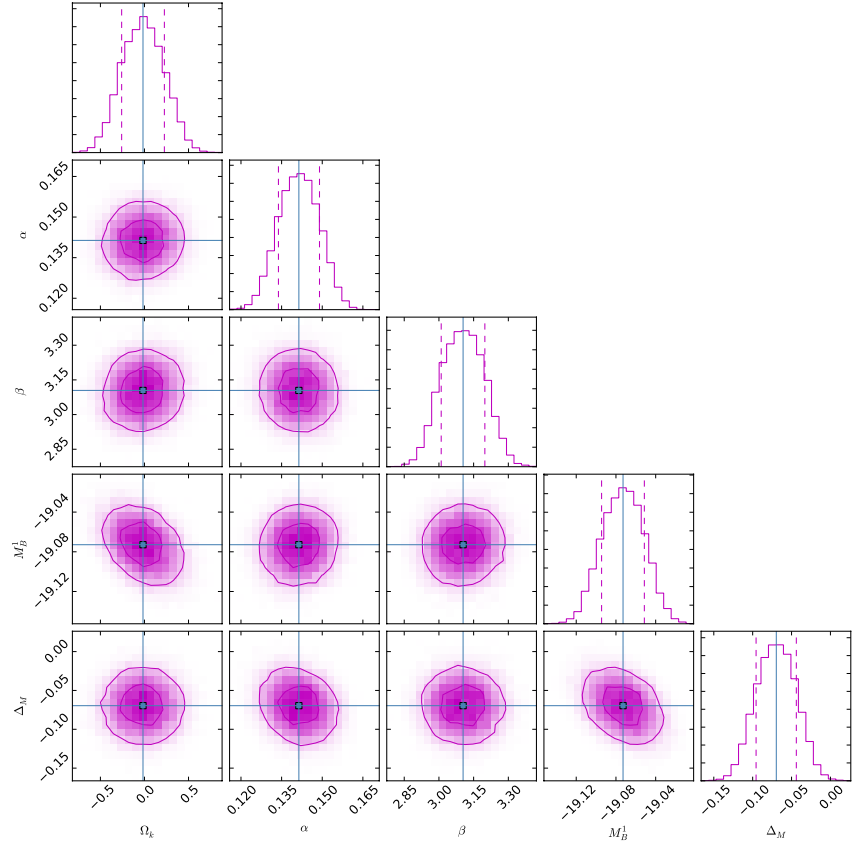} \\{(a) \;$H_{0}=69.6\pm0.7$ km $\rm s^{-1}$ $\rm Mpc^{-1}$}\\
\includegraphics[keepaspectratio,clip,width=0.65\textwidth]{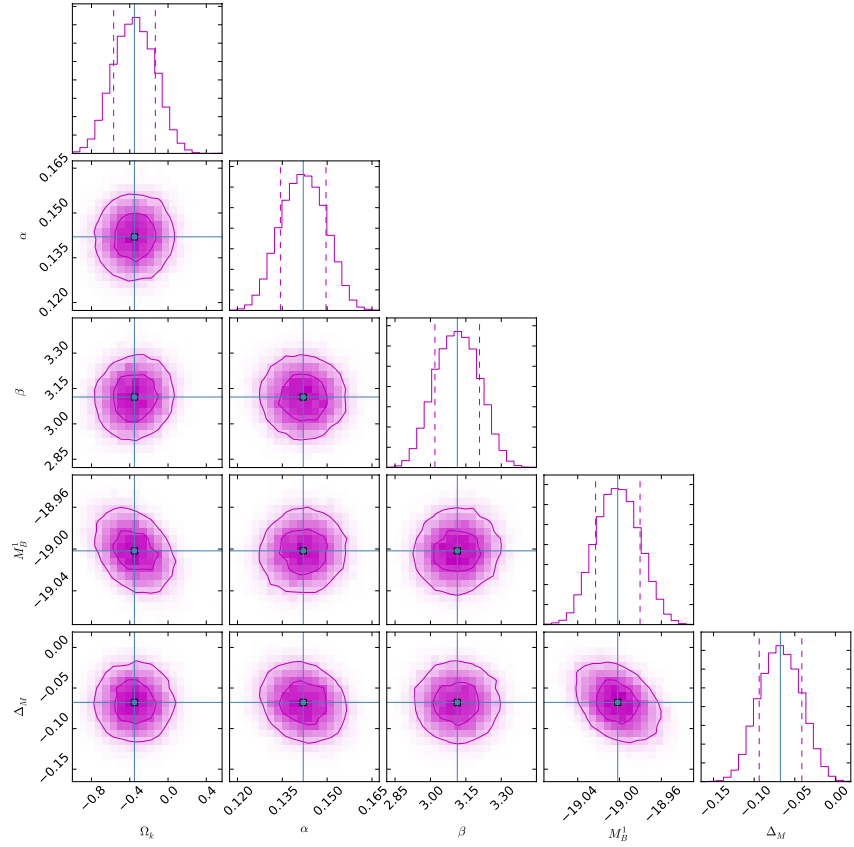} \\{(b) \;$H_{0}=73.24\pm1.74$ km $\rm s^{-1}$ $\rm Mpc^{-1}$}
\end{tabular}
\caption{Same as Figure~\ref{f3}, except now using the total $H(z)$ + SNe Ia data.}
\label{f4}
\end{figure*}

We show the constraints for the total $H(z)$ + JLA SNe Ia data in Figure~\ref{f4}.
The best-fit values corresponding to the priors of $H_{0}=69.6\pm0.7$ km $\rm s^{-1}$ $\rm Mpc^{-1}$
and $H_{0}=73.24\pm1.74$ km $\rm s^{-1}$ $\rm Mpc^{-1}$ are $\Omega_{k}=-0.02\pm0.24(1\sigma)\pm0.47(2\sigma)$ and
$\Omega_{k}=-0.35\pm0.22(1\sigma)\pm0.43(2\sigma)$, respectively (see Table~\ref{table2}).
Evidence also shows that no significant deviation from flatness is found. The best-fit $\Omega_{k}$
is in full agreement with zero spatial curvature at the $1.6\sigma$ confidence level, regardless of
which prior of $H_{0}$ is adopted. However, the influence of $H_{0}$ in this type of data is still exist.
That is, the prior of $H_{0}=73.24\pm1.74$ km $\rm s^{-1}$ $\rm Mpc^{-1}$ leads to a slightly bigger
deviation from the flat Universe. The comparison between Figure~\ref{f3} and Figure~\ref{f4}
(see also Table~\ref{table2}) shows that the best-fit results are more or less the same for both
the CC $H(z)$ + SNe Ia and total $H(z)$ + SNe Ia data, for the same prior of $H_{0}$.

\section{Summary and discussion}\label{sec:summary}

\citet{2007JCAP...08..011C,2008PhRvL.101a1301C} have proposed a model-independent method for
measuring the cosmic curvature. Using this method, several studies have been done. However,
we find that the luminosity distances of SNe Ia used in past works were obtained directly from
Hubble diagrams where the SN light-curve fitting parameters were inferred from global-fitting
in the context of a cosmological model. In contrary to previous studies, we keep the
light-curve fitting parameters free to investigate whether the curvature parameter has a
dependence on them. On the other hand, the estimation of the derivative function of comoving
distance $D(z)$ in the method of \citet{2007JCAP...08..011C,2008PhRvL.101a1301C} will introduce
a large uncertainty \citep{2016ApJ...828...85Y}.

In this work, we propose an improved model-independent method to
test the cosmic curvature. The main idea of our method is to compare two kinds of
distance moduli. One distance modulus $\mu_{H}(\Omega_{k})$ is constructed from
the $H(z)$ data, which is susceptible to the curvature parameter $\Omega_{k}$.
Based on the measurements of $H(z)$, we use the GP method to reconstruct
the $E(z)$ function and use Equation~(\ref{eq:dp}) to derive the proper distance
function $d_{P}(z)$. Using the reconstructed $d_{P}(z)$ function, the luminosity
distance $D^{H}_{L}(\Omega_{k})$ and the corresponding distance modulus
$\mu_{H}(\Omega_{k})$ from the $H(z)$ data can be further calculated at a certain $z$.
The other distance modulus $\mu_{\rm SN}(\alpha,\;\beta,\;M^{1}_{B},\;\Delta_{M})$
is from the SNe Ia data, which is inferred directly from the observed SN light-curve
(i.e., the original data $m^{\star}_{B}$, $X_{1}$, $\mathcal{C}$), but with some nuisance parameters
$(\alpha,\;\beta,\;M^{1}_{B},\;\Delta_{M})$.

Our model-independent analysis suggests that the best-fit curvature parameter is constrained to be
$\Omega_{k}=-0.02\pm0.24$, which is in good agreement with a flat Universe. We also considered
the impact of Hubble constant $H_{0}$ on the constraints, finding that different
$H_{0}$ priors can affect the measurements of $\Omega_{k}$ in some degree,
the prior of $H_{0}=73.24\pm1.74$ km $\rm s^{-1}$ $\rm Mpc^{-1}$ leads to a slightly bigger
deviation from the zero cosmic curvature than the other one of
$H_{0}=69.6\pm0.7$ km $\rm s^{-1}$ $\rm Mpc^{-1}$. In addition, we also compared
the constraints from different $H(z)$ samples: (i) the only CC $H(z)$ data; and (ii) the total $H(z)$ data.
We found that the optimized curvature parameters change quantitatively, though the qualitative results
and conclusions remain the same, independent of which kind of the sample is used.

In JLA \citep{2014A&A...568A..22B}, the SN nuisance parameters $(\alpha,\;\beta,\;M^{1}_{B},\;\Delta_{M})$
are derived from a fit to the flat $\Lambda$CDM model. In other words, \citet{2014A&A...568A..22B} compared
$\mu_{\rm SN}(\alpha,\;\beta,\;M^{1}_{B},\;\Delta_{M})$ with $\mu^{\rm \Lambda CDM}_{\rm th}(\Omega_{m})$
to find the best-fit cosmological parameters and nuisance parameters, which were
$\Omega_{m}=0.30\pm0.03$, $\alpha=0.14\pm0.01$, $\beta=3.10\pm0.08$, $M^{1}_{B}=-19.05\pm0.02$,
and $\Delta_{M}=-0.07\pm0.02$. In our analysis, we adopt the constructed $\mu_{H}(\Omega_{k})$
from the $H(z)$ data, instead of the cosmology-dependent $\mu^{\rm \Lambda CDM}_{\rm th}(\Omega_{m})$,
and then derive the best-fit curvature parameter and nuisance parameters by comparing $\mu_{H}(\Omega_{k})$
with $\mu_{\rm SN}(\alpha,\;\beta,\;M^{1}_{B},\;\Delta_{M})$. We find that our constraints on
the nuisance parameters (see Table~\ref{table2}) are very similar to those results of
\citet{2014A&A...568A..22B},\footnote{Note that the only one parameter that different is
$M^{1}_{B}$. Since it is degenerate with $H_{0}$, it has to change if $H_{0}$ changes.}
not only attesting to the reliability of our calculation, but also confirming that the curvature parameter
is independent of the nuisance parameters.

\begin{figure*}
\centerline{\includegraphics[keepaspectratio,clip,width=0.5\textwidth]{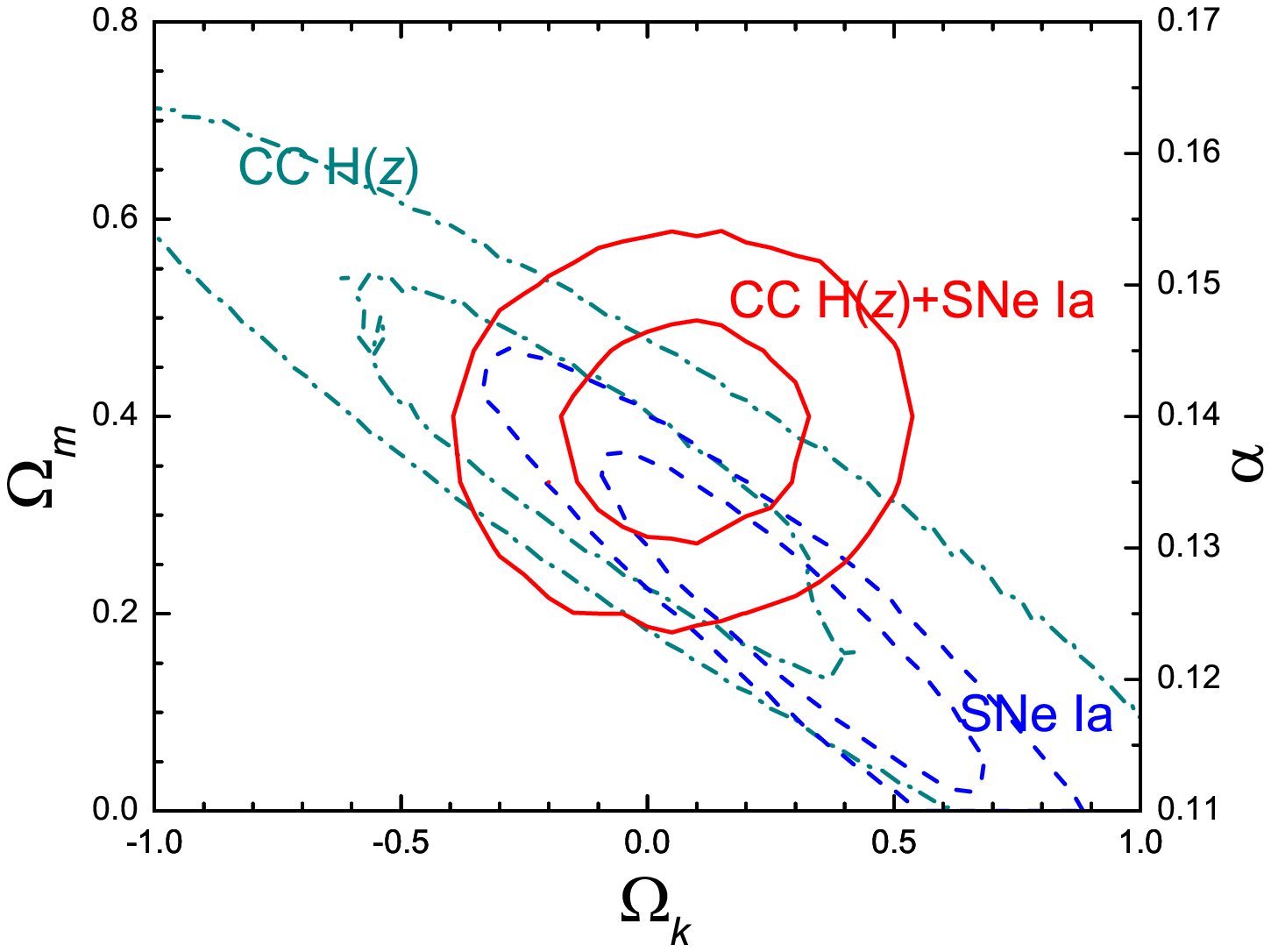}}
\vskip-0.1in
\caption{$1\sigma$ and $2\sigma$ constraint contours of $(\Omega_{k},\;\Omega_{m})$
in the $\Lambda$CDM model determined with the conventional method for CC $H(z)$ (dark cyan dash-dotted lines)
and SNe Ia (blue dashed lines), respectively. The red solid contours correspond to the confidence levels of
$(\Omega_{k},\;\alpha)$ for the CC $H(z)$ + SNe Ia data, obtained from the cosmological model-independent method (i.e., the GP method).}
\label{f5}
\end{figure*}

To check the validity and efficiency of our new method, we also run the more conventional
(non-Gaussian processes) method and just leave the curvature parameter free to see what value we get and
if it is different. Following the conventional method, we allow $\Omega_{k}$ to be free along with the matter energy density
$\Omega_{m}$ in the $\Lambda$CDM model, and compare $\mu_{\rm SN}(\alpha,\;\beta,\;M^{1}_{B},\;\Delta_{M})$ with
$\mu^{\rm \Lambda CDM}_{\rm th}(\Omega_{m},\;\Omega_{k})$ (or compare $H_{\rm obs}(z)$ with $H^{\rm \Lambda CDM}_{\rm th}(z;\;\Omega_{m},\;\Omega_{k})$).
In Figure~\ref{f5}, we display the confidence regions of $(\Omega_{k},\;\Omega_{m})$
in the $\Lambda$CDM model determined with the conventional method for CC $H(z)$ (dark cyan dash-dotted lines) and SNe Ia (blue dashed lines), respectively.
The contours show that at the $1\sigma$ confidence level, the best-fits are
$(\Omega_{k}=0.30\pm0.39,\;\Omega_{m}=0.20\pm0.17)$ for SNe Ia and $(\Omega_{k}=0.03^{+0.64}_{-0.55},\;\Omega_{m}=0.33^{+0.19}_{-0.21})$ for CC $H(z)$.
The corresponding contours of $(\Omega_{k},\;\alpha)$ from our GP method for the CC $H(z)$ + SNe Ia data (red solid lines) are also shown in Figure~\ref{f5} for comparisons.
One can see that the determined $\Omega_{k}$ from the conventional method
are also consistent with a flat Universe within error limits. But, the errors on these measured
$\Omega_{k}$ are at the levels of $\sigma_{\Omega_{k}}\simeq0.39$ and $\sigma_{\Omega_{k}}\simeq0.64$, which
are not as good as that of our GP method $(\sigma_{\Omega_{k}}\simeq0.22)$.
What's more, our constraint on $\Omega_{k}$ with the GP method is more robust and more widely applicable as it does not depend on the cosmological model.
If in the future the quality of observational data are much improved, the prospects
for constraining the cosmic curvature with this method will be very promising.

\acknowledgments
We are very grateful to the anonymous referee for providing a thoughtful
review and making several important suggestions that have improved the
manuscript significantly. We also acknowledge Gabriel R. Bengochea for useful communications.
This work is partially supported by the National Basic Research Program (``973'' Program)
of China (Grant No. 2014CB845800), the National Natural Science Foundation of China
(Grant Nos. 11673068 and 11603076), the Youth Innovation Promotion
Association (2011231 and 2017366), the Key Research Program of Frontier Sciences (QYZDB-SSW-SYS005),
the Strategic Priority Research Program ``Multi-waveband gravitational wave Universe''
(Grant No. XDB23000000) of the Chinese Academy of Sciences, the Natural Science Foundation
of Jiangsu Province (Grant No. BK20161096), and the Guangxi Key Laboratory for Relativistic Astrophysics.


\end{document}